\documentclass[]{llncs} 
\usepackage{graphicx} 
\usepackage{amssymb,algorithmic}

\begin{document}

\newtheorem{algorithm}{Algorithm}

\title{Parking functions, labeled trees\\ and DCJ sorting scenarios$^*$}

\author{A\"ida Ouangraoua\inst{1,2}
\and Anne Bergeron\inst{2}}

\institute{ Department of Mathematics, Simon Fraser University,
 Burnaby (BC), Canada \email{aouangra@sfu.ca} \and Lacim, 
Universit\'e du Qu\'ebec \`a Montr\'eal, Montr\'eal (QC), Canada.
\email{bergeron.anne@uqam.ca}}

\maketitle

\let\thefootnote\relax\footnotetext{$^*$Date: January 10, 2009.\\  Manuscript under preparation.} 
%
%
\begin{abstract}
In genome rearrangement theory, one of the elusive questions
raised in recent years is the enumeration of rearrangement scenarios
between two genomes. This problem is related to the uniform generation
of rearrangement  scenarios, and the derivation of tests of statistical
significance of the properties of these scenarios. Here we give an exact formula for the
number of double-cut-and-join (DCJ) rearrangement scenarios of co-tailed genomes.
We also construct effective bijections between the set of scenarios that sort a cycle and 
well studied combinatorial objects such as parking functions and labeled trees.
\end{abstract}

%
%

\section{Introduction}

Sorting genomes can be succinctly described as finding sequences 
of rearrangement operations that transform a genome into another.
The allowed rearrangement operations are fixed, and the sequences
of operations, called {\em sorting scenarios}, are ideally of minimal length.
Given two genomes, the number of different sorting scenarios between them
is typically huge -- we mean HUGE -- and very few analytical tools
are available to explore these sets.

In this paper, we give the first exact results on the enumeration
and representation of sorting scenarios in terms of well-known 
combinatorial objects. We prove that sorting scenarios using
DCJ operations on {\em co-tailed} genomes can be represented by 
{\em parking functions} and {\em labeled trees}. This surprising connection
yields immediate results on the uniform generation of scenarios 
\cite{Ajana-2002,Miklos-2006,Siepel-2002}, promises tools for sampling processes 
\cite{Braga-2008,Larget-2002,Mikos-Hein-2004} and the development 
of statistical significant tests \cite{Dalevi-2002,McLysaght-2000,Sankoff-2004,Xu-2008},
and offers a wealth of alternate representations to explore the 
properties of rearrangement scenarios, such as commutation \cite{Bergeron-2002,Swenson2009}, 
structure conservation \cite{Berard-2007,Diekmann-2007}, breakpoint reuse 
\cite{pevzner:tesler2003,sankoff:trinh2005} or cycle length \cite{Xu-2007}.

This research was initiated while we were trying to understand {\em commuting}
operations in a general context. In the case of genomes consisting of single chromosomes, 
rearrangement operations are often modeled as {\em inversions}, which can be represented 
by intervals of the set $\{1, 2, \ldots, n\}$. Commutation properties are described 
by using overlap relations on the corresponding sets, and a major tool to understand 
sorting scenarios are {\em overlap graphs}, whose vertices represent single rearrangement operations,
and whose edges model the interactions between the operations.
Unfortunately, overlap graphs do not upgrade easily to genomes with multiple chromosomes,
see, for example, \cite{Ozery-Shamir-2007}, where a generalization is given for a restricted set of 
operations.

We got significant insights when we switched our focus from single rearrangement operations
to complete sorting scenarios. This apparently more complex formulation offers
the possibility to capture complete scenarios of length $d$ as simple combinatorial objects,
such as sequences of integer of length $d$, or trees with $d$ vertices. It also gives alternate
representations of sorting scenarios, using {\em non-crossing} partitions, that facilitate the study of
commuting operations and structure conservation.

In Section~\ref{Representation}, we first show that sorting a cycle in 
the adjacency graph of two genomes with DCJ 
rearrangement operations is equivalent to refining
non-crossing partitions. This observation, together with a result
by Richard Stanley \cite{Stanley-1997}, gives the  existence of bijections between 
sorting scenarios of a cycle and parking functions or labeled trees.
We give explicit bijections for both in Sections~\ref{ParkingFunctions} 
and \ref{LabeledTrees}. We conclude in Section~\ref{Discussion} with remarks
on the usefulness of these representations, on the algorithmic complexity
of switching between representations, and on generalizations to genomes
that are not necessarily co-tailed.

%
%

\section{Preliminaries}

Genomes are compared by identifying homologous segments along
their DNA sequences, called {\em blocks}.  These blocks can be
relatively small, such as gene coding sequences, or very large fragments
of chromosomes. The order and orientation of the blocks may vary in
different genomes. Here we assume that the two genomes
consist of either circular chromosomes, or {\em co-tailed} linear chromosomes. 
For example, consider the following two genomes, each consisting of two linear chromosomes:

\begin{itemize}
\item[] Genome $A$:
  ~~~~  ({\em a -f  -b~ e -d}) ({\em -c~ g})
\item[] Genome $B$:
  ~~~~  ({\em a~ b~ c}) ({\em d~ e~ f ~g})
\end{itemize}

The set of {\em tails} of a linear chromosome $(x_1 \ldots x_m)$ is $\{x_1, -x_m\}$,
and two genomes are {\em co-tailed} if the union of their sets of tails are the same. 
This is the case for genomes $A$ and $B$ above, since the the union of their sets
of tails is $\{a, -c, d, -g\}$.

An {\em adjacency} in a genome is a sequence of two consecutive blocks. For
example, in the above genomes, ({\em e  -d}) is an adjacency of genome $A$,
and ({\em a~ b}) is an adjacency of genome $B$.
Since a whole chromosome can be flipped, we always have
$( x~ y) = ({-y}~{-x})$.

\begin{figure}[h]
\centering
\includegraphics[width=12cm]{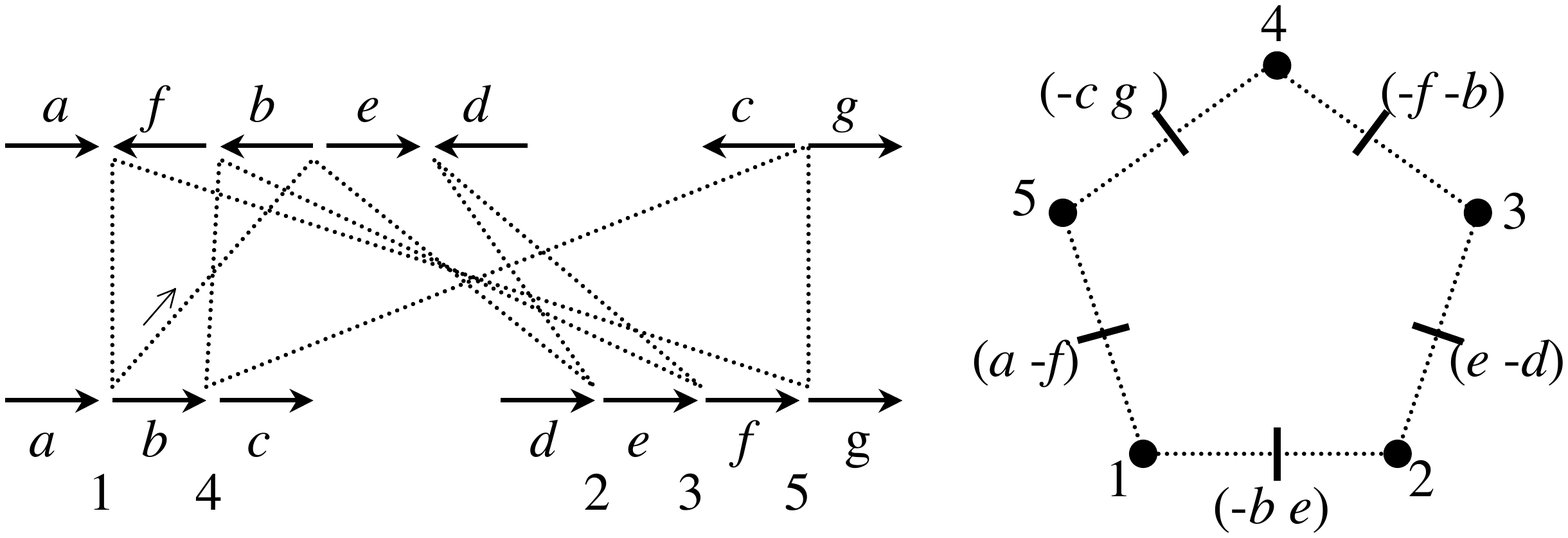}
\caption{At the left, the adjacency graph of genome $A$ =  ({\em a -f  -b~ e -d}) ({\em -c~ g}) and
genome $ B$ = ({\em a~ b~ c})({\em d~ e~ f ~g}) is represented by dotted lines. 
The sign of a block is represented by the orientation of the corresponding arrow.
If the -- single -- cycle is traversed starting with an arbitrary adjacency of genome $B$, 
here ({\em a~ b}), in the direction of the small arrow, 
then the 5 adjacencies of genome $B$ will be visited in the order
indicated by the numbers 1 to 5. At the right, the cycle has been spread out,
showing that any DCJ operation acting on two adjacencies of genome $A$ that splits the cycle can be represented 
by two cuts on the cycle $(12345)$.
 }
\label{Graphe}
\end{figure}

The {\em adjacency graph} of two genomes $A$ and $B$ is a graph whose
vertices are the adjacencies of $A$ and $B$, and such that for each block $y$ there is
an edge between adjacency $(y~z)$ in genome $A$ and $(y~z')$ in genome
$B$, and an edge between $(x~y)$ in genome $A$, and $(x'~y)$ in genome $B$.
See, for example, Figure~\ref{Graphe}.

Since each vertex has two incident edges, the adjacency graph
can be decomposed into connected components that are cycles. 
The graph of Figure~\ref{Graphe} has a single cycle of length 10.

A {\em double-cut-and-join} (DCJ) rearrangement operation \cite{Bergeron-2006,Yancopoulos-2005}
on genome $A$ acts on two adjacencies $(x~y)$ and
$(u~v)$ to produce either  $(x~v)$ and $(u~y)$, or $(x~{-u})$ and $({-y}~v)$.
In simpler words, a DCJ operation cuts the genome at two places, and
glues the part in a different order. 

The {\em distance} between
genomes $A$ and $B$ is the minimum number of DCJ operations needed to
rearrange -- or {\em sort} -- genome $A$ into genome $B$.
The DCJ distance is easily computed from the adjacency graph \cite{Bergeron-2006}.
For circular chromosomes or co-tailed genomes, the distance is given by:
$$ d(A, B) = N - (C + K)$$
where $N$ is the number of blocks, $C$ is the
number of cycles of the adjacency graph, and $K$ is the number of linear chromosomes in $A$. 
Note that $K$ is a constant for co-tailed genomes.
A rearrangement operation is {\em sorting} if it lowers the distance by 1, and a sequence of sorting operations
of length $d(A, B)$ is called a {\em parsimonious sorting scenario}. It is easy
to detect sorting operations since, by the distance formula, a sorting operation must increase by 1 the
number of cycles. 

A DCJ operation that acts on two cycles of the adjacency graph will merge
the two cycles, and can never be sorting. Thus the sorting operations act on a
single cycle, and split it into two cycles.  
The central question of this paper is to enumerate the set of
parsimonious sorting scenario. Since each cycle is sorted independently of the others,
the problem reduces to enumerating the sorting scenarios
of a cycle. Indeed, we have:

\begin{proposition}
\label{EnumerationSimple}
Given scenarios $S_1, \ldots, S_C$ of lengths $\ell_1, \ldots, \ell_C$ that sort
the $C$ cycles of an adjacency graph, these scenarios can be shuffled into a global scenario in
$$ \left( \begin{array}{c} \ell_1 + \ell_2 + \ldots + \ell_C \\ \ell_1,  \ell_2,  \ldots , \ell_C \end{array} \right) 
= \frac{(\ell_1 + \ell_2 + \ldots + \ell_C ) !}{\ell_1!  \ell_2!  \ldots  \ell_C !}$$
different ways.
\end{proposition} 
\begin{proof}
Since each cycle is sorted independently, the number of global scenarios is enumerated by 
counting the number of sequences
that contains $\ell_m$ occurrences of the symbol $S_m$, for $1 \leq m \leq C$, which is counted
by a classical formula. For each such
sequence, we obtain a scenario by replacing each symbol $S_m$ by the appropriate
operation on cycle number $m$.
\end{proof}

%
%

\section{Representation of scenarios as sequences of fissions}
\label{Representation}

A cycle of length $2n$ of the adjacency graph alternates between adjacencies of genome A and genome B.
Given a cycle, suppose that the adjacencies of genome $B$ are labeled by integers from $1$ to $n$ in the order
they appear along the cycle, starting with an arbitrary adjacency (see Fig.~\ref{Graphe}). Then any DCJ operation
that splits this cycle can be represented by a {\em fission} of the cycle  $(1 2 3 \ldots n)$,
as  $$(1 2 3 \ldots p \|q \ldots t\|u \ldots n)$$
yielding the two cycles:
$$(1 2 3 \ldots pu \ldots n) \mbox{ and }(q \ldots t).$$
We will always write cycles beginning with their smallest  element. Fissions applied to a cycle
whose elements are in increasing order always yield cycles whose elements are in increasing order. A fission is characterized
by two cuts, each described by the element at the left of the cut. The smallest one, $p$ in the above
example, will be called the {\em base} of the fission, and the largest one, $t$ in the the above example, is
called the {\em top} of the fission.
The integer at the right of the first cut, $q$ in the example, is called the {\em partner}
of the base. 

In general, after the application of $k$ fissions on  $(1 2 3 \ldots n)$,  the resulting set of cycles
will contain $k+1$ elements. The structure of these cycles form a {\em non-crossing
partition} of the initial cycle  $(1 2 3 \ldots n)$. Namely, we have the following result, which is easily
shown by induction on $k$:

\begin{proposition} \label{NonCrossingProperty}
Let $k \leq n-1$ fissions be applied on
the cycle  $(1 2 3 \ldots n)$, then the $k+1$ resulting cycles have the following
properties:\\

\noindent 1)  The elements of each cycle are in increasing order, up to cyclical reordering.\\

\noindent  2) [Non-crossing property] If  $(c \ldots d)$ and $(e \ldots f)$ are two cycles with $c < e$, then
either $d <  e$,  or $c < e\leq  f < d$.\\

\noindent 3) Each successive fission refines the partition of  $(1 2 3 \ldots n)$ defined by the cycles.

\end{proposition}

A sorting scenario of a cycle of length $2n$ of the adjacency graph can thus be represented by a sequence
of $n-1$ fissions on the cycle $(1 2 3 \ldots n)$, called a {\em fission scenario},
and the resulting set of cycles will have the structure  $(1)( 2)( 3) \ldots (n)$.
For example, here is a possible fission scenario of $(1 2 3 4 5 6 7 8 9)$, where the bases 
of the fissions have been underlined:

\begin{center}
$(1 2 3  \underline{4}  \| 5 \| 6 7 8 9) \rightarrow (1 2 3 4 6 7 8 9) (5)$

$(1 2 3 4 6  7  \underline{8} \|9\|) (5) \rightarrow  (1 2 3 4 6  7 8 ) (5) (9)$

$( \underline{1} \| 2 3 4 6  7 8 \|) (5) (9)  \rightarrow  (1) ( 2 3 4 6  7 8 ) (5) (9)  $

$ (1) (  \underline{2} \| 3 4 6 \|  7 8) (5) (9)   \rightarrow    (1) ( 2  7 8) ( 3 4 6 ) (5) (9) $

$   (1) (  \underline{2} \| 7 \| 8) ( 3 4 6 ) (5) (9)  \rightarrow   (1) ( 2  8) ( 3 4 6 ) (5) (7) (9) $

$  (1) ( 2  8) (  \underline{3} \| 4 6 \|) (5) (7) (9)   \rightarrow  (1) ( 2  8) ( 3) (4 6 ) (5) (7) (9)   $

$  (1) (  \underline{2} \| 8\|) ( 3) (4 6 ) (5) (7) (9)   \rightarrow  (1) (2) ( 3) (4 6 ) (5) (7) (8) (9)   $

$  (1) (2) ( 3) ( \underline{4} \| 6 \|) (5) (7) (8) (9)    \rightarrow  (1) (2) ( 3) (4) (5) (6) (7) (8) (9)   $
\end{center}

Scenarios such as the one above have interesting combinatorial features when all the operations
are considered globally, and we will use them extensively in the sequel. A first important remark is
that the smallest element of the cycle is always `linked' to the greatest element through a chain of partners.
For example, the last partner of element 1 is element 2, the last partner of element 2 is element 8,
and the last partner of element 8 is element 9. We will see that this is always the case, even  when the 
order of the corresponding fissions is arbitrary with respect to the scenario. The following definition
captures this idea of chain of partners.

\begin{definition}
Consider a scenario $S$ of fissions that transform a cycle $(c \ldots d)$ into
cycles of length 1. For each element $p$ in $(c \ldots d)$, if $p$ is the base of one or more of the 
fissions of $S$, let $q$ be the last
partner of $p$, then define recursively  $$Sup_S(p) = Sup_S(q),$$ otherwise, $Sup_S(p) = p$.
\end{definition}

In order to see that $Sup(p)$ is well defined, first note that the successive partners of a given base $p$
are always in increasing order, and greater than $p$. Moreover, the last element of a cycle $(c \ldots d)$
is never the base of a fission. For example, in the above scenario, we would have $Sup_S(1) = Sup_S(2) = Sup_S(8) = Sup_S(9) = 9$.

The following lemma is the key to most of the results that follow:

\begin{lemma}
\label{lemma_sup}
Consider a scenario $S$ of fissions that transform a cycle $(c \ldots d)$ into
cycles of length 1, then $Sup_S(c) = d$.
\end{lemma}
\begin{proof}
If $c = d$, then the result is trivial. Suppose the result is true for cycles of length $ \leq n$, and
consider a cycle of length $n+1$. The first fission of $S$ will split the cycle  $(c \ldots d)$ in two
cycles of length $ \leq n$. If the two cycles are of the form $(c \ldots d)$ and $(c' \ldots d')$, then
$c'$ is, in the worst case, the first partner of $c$, and cannot be the last since $c \neq d$. 
Let $S'$ be the subset of $S$ that  transform the shorter cycle $(c \ldots d)$ into
cycles of length 1. By the induction hypothesis, 
$Sup_{S'}(c) = d$, but  $Sup_S(c) =  Sup_{S'}(c)$ since the last partner of $c$ is not in $(c' \ldots d')$.

If the two cycles are of the form $(c \ldots d')$ and $(c' \ldots d)$, consider $S_1$ the subset of $S$ that  transform 
the cycle $(c \ldots d')$ into cycles of length 1, and $S_2$  the subset of $S$ that  transform 
the cycle $(c' \ldots d)$ into cycles of length 1. We have, by the induction hypothesis,  $Sup_{S_1}(c) = d'$ 
and  $Sup_{S_2}(c') = d$, implying $Sup_S(c) = Sup_S(d')$ and  $Sup_S(c') = d$. However, $c'$ is the last partner of $d'$, 
thus $Sup_S(c) = Sup_S(d') = Sup_S(c') = d$.
\end{proof}

%
%

\section{Fission scenarios and parking functions}
\label{ParkingFunctions}

In this section, we establish a bijection between fission scenarios
and {\em parking functions} of length $n-1$. This yields a very compact
representation of DCJ sorting scenarios of cycles of length $2n$ as sequences of $n-1$ integers.

A {\em parking function} is a sequence of integers 
$p_1 p_2 \ldots p_{n-1}$
such that if the sequence is sorted in non-decreasing order yielding
$p'_1 \leq  p'_2  \leq \ldots \leq p'_{n-1}$, then $p'_i \leq i$. These sequences were 
introduced by Konheim and Weiss \cite{Konheim-Weiss-1966}  in connection with hashing problems.
These combinatorial structure are well studied, and the number of different
parking functions of length $n-1$ is known to be  $n^{n-2}$.

Proposition~\ref{NonCrossingProperty} states that a fission scenario is a sequence
of successively refined non-crossing partitions of the cycle $(1 2 3 \ldots n)$. A result
by Stanley \cite{Stanley-1997} has the following immediate consequence:

\begin{theorem}
\label{Bijection}
There exists a bijection between fission scenarios of cycles
of the form $(1 2 3 \ldots n)$ and {\em parking functions} of length $n-1$. 
\end{theorem}

Fortunately, in our context, the bijection is very simple: we
list the bases of the fissions of the scenario. For example,
the parking function associated to the example of Section~\ref{Representation} is 48122324.
In general, we have:

\begin{proposition}
The sequence of bases of a fission scenario on the cycle $(1 2 3 \ldots n)$ is a parking function of length $n-1$.
\end{proposition}
\begin{proof}
Let  $p_1 p_2 \ldots p_{n-1}$ be the sequence of bases of a fission scenario 
and let  $p'_1 p'_2 \ldots p'_{n-1}$ be the corresponding sequence sorted in non-decreasing order. 
Suppose that there exists a number $i$ such that  $p'_i > i$, then there are at least $n-i$ fissions
in the scenario with base $p \geq i+1$. These bases can be associated to at most $n-i-1$ partners 
in the set $\{i+2,i+3,i+4, \ldots ,n\}$ because a base is always smaller than its partner, but this 
is impossible because each integer is used at most once as a partner in a fission scenario. 
\end{proof}

In order to reconstruct a fission scenario from a parking function, we first note that a fission with base $p_i$ and partner $q_i$
creates a cycle whose smallest element is $q_i$, thus 
each integer in the set $\{2, 3, \ldots , n \}$ appears exactly once as a partner in a fission scenario.

Given a parking function $p_1 p_2 \ldots p_{n-1}$, we must first assign to each base $p_i$ a 
unique partner $q_i$ in the set $\{2, 3, \ldots , n \}$. 
By Lemma~\ref{lemma_sup}, we can then determine the top $t_i$ of fission $i$, since 
the set of fissions from $i+1$ to $n -1$ contains a sorting scenario of the cycle $(q_i \ldots t_i)$.  
Algorithm 1 details the procedure.\\

\noindent {\bf Algorithm 1 [Parking functions to fission scenarios]}

\noindent Input: a parking function  $p_1 p_2 \ldots p_{n-1}$.

\noindent Output: a fission scenario $(p_1, t_1), \ldots,  (p_{n-1}, t_{n-1})$.\\

{\em $Q \leftarrow \{2, 3, \ldots , n \}$

For $p$ from $n-1$ to 1 do:

~~~For each successive occurrence $p_i$ of $p$ in the sequence  $p_1 p_2 \ldots p_{n-1}$ do:

~~~~~~$q_i \leftarrow $ The smallest element of $Q$ greater than $p_i$
    
~~~~~~$Q \leftarrow Q \setminus \{q_i \}$

$S  \leftarrow \{(p_1,q_1),(p_2,q_2) \ldots, (p_{n-1},q_{n-1})\}$ 

For  $i$ from $1$ to ${n- 1}$ do:

~~~ $S  \leftarrow S \setminus \{(p_i,q_i) \}$

~~~ $t_i  \leftarrow Sup_S(q_i)$}\\

For example, using the parking function 48122324 and the set of partners $\{2, 3, \ldots , 9 \}$,
we would get the pairings, starting from base 8 down to base 1:
$$ \left( \begin{array}{ccccccccc}
p_i: & 4 & 8 & 1 & 2 & 2 & 3 & 2 & 4 \\
q_i: & 5 & 9 & 2 & 3 & 7 & 4 & 8 & 6
\end{array} \right)$$

Finally, in order to recover the second cut of each fission, we compute the values $t_i$:
$$ \left( \begin{array}{ccccccccc}
p_i: & 4 & 8 & 1 & 2 & 2 & 3 & 2 & 4 \\
t_i: & 5 & 9 & 8  & 6 & 7 & 6 & 8 & 6
\end{array} \right)$$
For example, in order to compute $t_4$, then $S = \{(2,7), (3,4), (2,8), (4,6)\}$,  and $Sup_{S}(3) = 
Sup_S(4) = Sup_S(6) = 6.$

Since we know, by Theorem~\ref{Bijection}, that fissions scenarios are in bijection with parking
functions, it is sufficient to show that  Algorithm~1 recovers a given scenario in order
to prove that it is an effective bijection.

\begin{proposition}
Given a fission scenario of a cycle of the form $(1 2 3 \ldots n)$,
let $(p_i, q_i, t_i)$ be the base, partner and top of fission $i$. Algorithm 1 recovers uniquely $t_i$ 
from the parking function $p_1 p_2 \ldots p_{n-1}$.
\end{proposition}
\begin{proof}
By Lemma~\ref{lemma_sup} , we only need to show that 
Algorithm~1 recovers uniquely the partner $q_i$ of each base $p_i$.
Let $p$ be the largest base, and suppose that $p$ has $j$ partners,
then the original cycle must contain at least the elements:
$$(\ldots p~p+1 \ldots p+j \ldots ).$$
We will show that $p+1 \ldots p+j$ must be the $j$ partners of $p$. If it was not the case,
at least one of the $j$ adjacencies in the sequence $ p~p+1 \ldots p+j$ must be cut 
in a fission whose base is smaller than $p$, since $p$ is the largest base, and this would 
violate the non-crossing property of Proposition~\ref{NonCrossingProperty}. Thus 
Algorithm~1 correctly and uniquely assigns the partners of the largest base.
Suppose now that Algorithm~1 has correctly and uniquely assigned the partners of all
bases greater than $p$. The same argument shows that the successive partners
of $p$ must be the smallest available partners greater than $p$.
\end{proof}

Summarizing the results so far, we have:
\begin{theorem}
If the adjacency graph of two co-tailed genomes has $C$ cycles of length $2(\ell_1 + 1), \ldots, 2(\ell_C + 1)$,
then the number of sorting scenarios is given by:
$$\frac{(\ell_1 + \ell_2 + \ldots + \ell_C ) !}{\ell_1!  \ell_2!  \ldots  \ell_C!} * {(\ell_1 + 1)}^{\ell_1 -1} * \ldots * {(\ell_C +1) }^{\ell_C-1}. $$
Each sub-scenario that sort a cycle of length  $2(\ell_m + 1)$  can be represented by a parking function 
of length $\ell_m $.
\end{theorem}
\begin{proof}Sorting a cycle of length $2(\ell_m + 1)$ can be simulated by fissions of the cycle $(12 \ldots \ell_m + 1)$,
which can be represented by parking functions of length $\ell_m$. The number of different parking functions of length $\ell_m$
is given by ${(\ell_m + 1)}^{\ell_m -1}$. Applying Proposition~\ref{EnumerationSimple} yields the enumeration formula.
\end{proof}

%
%

\section{Fission scenarios and labeled trees}
\label{LabeledTrees}

Theorem~\ref{Bijection} implies that it is possible to construct bijections between fission scenarios 
and objects that are enumerated by parking functions. This is notably the case of 
 {\em labeled tree} on $n$ vertices. These are trees with $n$ vertices 
in which each vertex is given a unique label in the set $\{0,1, \ldots, n-1\}$.
In this section, we construct an explicit bijection between these trees and fission scenarios
of cycles of the form $(1 2 3 \ldots n)$.

\begin{definition}
Given a fission scenario $S$ of a cycle of the form $(1 2 3 \ldots n)$,
let $(p_i, q_i)$ be the base and partner of fission $i$. 

The graph $T_S$ is a graph whose nodes are labeled by $\{0,1, \ldots, n-1\}$,  
with an edge between $i$ and $j$, if $p_i = q_j$, and an edge between $0$ and  $i$, if $p_i = 1$.
\end{definition}

\begin{figure}[h]
\centering
\includegraphics[width=12cm]{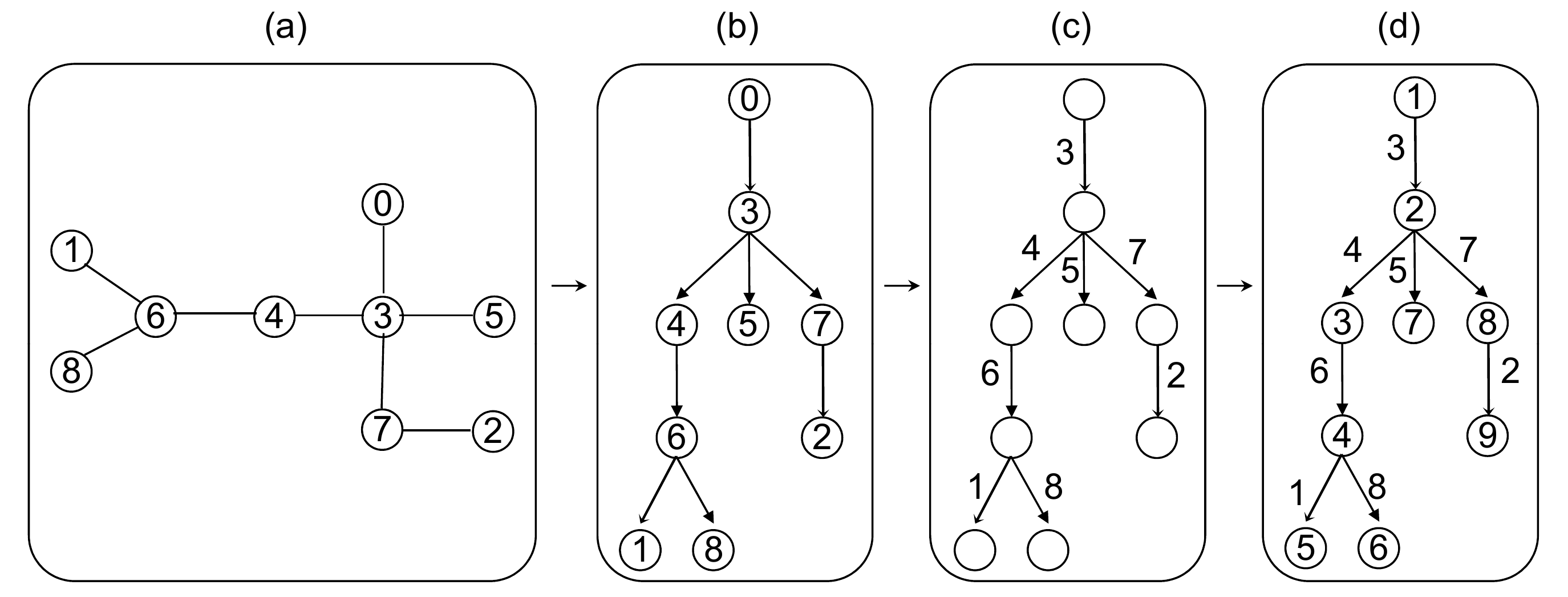}
\caption{Construction of a fission scenario. 
(a) The unrooted tree $T_S$. 
(b) The tree is rooted at vertex 0, and the children of each node are ordered.
(c) The labels of the nodes are lifted to their incoming edges.
(d) The nodes are labeled in prefix order from 1 to 9.
The order of the fissions are read on the edges, the source of an edge
represent the base $p$ of the fission and its target is the partner. 
For example, fission \#1 has base 4, with partner 5.  }
\label{tree-construction}
\end{figure}

In the running example, the corresponding graph is depicted in Figure \ref{tree-construction} (a). We have:
\begin{proposition}
The graph $T_S$ is a labeled tree on $n$ vertices.
\end{proposition}
\begin{proof}
By construction, the graph has $n$ vertices labeled by $\{0,1, \ldots, n-1\}$. In order to show
that it is a tree, we will show that the graph has $n-1$ edges and that it is connected. 
Since each integer in the set $\{2,3, \ldots, n\}$ is partner of one and only one fission
in $S$ and $S$ contains $n-1$ fissions, $T_S$ has exactly $n-1$ edges. Moreover, by construction,
there is a path between each vertex $i\neq 0$ and $0$ in $T_S$, thus $T_S$ is connected. 
\end{proof}

Before showing that the construction of $T_S$ yields an effective bijection, we
detail how to recover a fission scenario from a tree. \\

\noindent {\bf Algorithm 2 [Labeled trees to fission scenarios]}

\noindent Input: a labeled tree $T$ on $n$ vertices.

\noindent Output: a fission scenario $(p_1, t_1), \ldots,  (p_{n-1}, t_{n-1})$.\\

{\em Root the tree at vertex 0.

Put the children of each node in increasing order from left to right.

Label the unique incoming edge of a node with the label of the node.

Relabel the nodes from $1$ to $n$ with a prefix traversal of the tree.

For $i$ from $1$ to ${n-1}$ do:

~~~ $p_i \leftarrow $ The label of the source $p$ of edge $i$. 

~~~ $t_i  \leftarrow$ The greatest label of the subtree rooted by edge $i$.

~~~ Remove edge $i$ from $T$}\\

The following proposition states that the construction of the associated tree $T_S$ is
injective, thus providing a bijection between fission scenarios and trees.

\begin{proposition}
The trees associated to different fission scenarios are different.
\end{proposition}
\begin{proof}
Suppose that two different scenarios $S_1$ and $S_2$ yield the same tree $T$. Then, 
by construction, if $T$ is rooted in $0$, for each directed edge from $j$ to $i$ in $T$, 
if $j=0$ then $p_i = 1$ otherwise $p_i = q_j$. Moreover, in a fission scenario, 
if fission $i$ is the first operation having base $p_i$, then its partner is
$q_i = p_i+1$, otherwise the non-crossing property of Proposition~\ref{NonCrossingProperty}
would be violated. So, using these two properties, the sequences of bases and partners 
of the fissions in the two scenarios can be uniquely recovered from $T$, and thus 
$S_1$ and $S_2$ would correspond to the same parking function. 
\end{proof} 

The tree representation offers another interesting view of the sorting procedure. Indeed,
sorting can be done directly on the tree by successively erasing the edges from 1 to $n-1$.
This progressively disconnects the tree, and the resulting connected components correspond
precisely to the intermediate cycles obtained during the sorting procedure.

\begin{figure}[h]
\centering
\includegraphics[width=12cm]{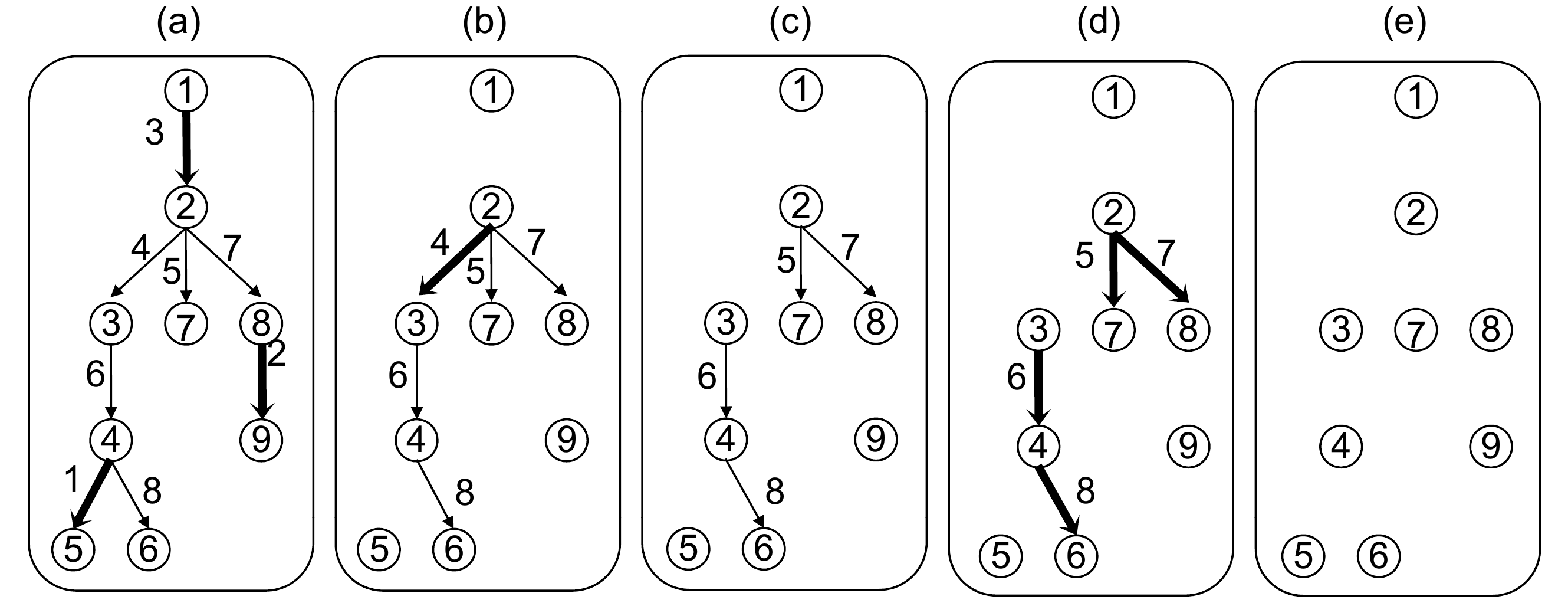}
\caption{Sorting directly on a tree: erasing successively  the edges 1 to $n$ 
simulates cycle fissions by creating intermediate forests.
In part (b), the fourth fission will split the tree corresponding to cycle $(234678)$
into two trees corresponding to the two cycles $(346)$ and $(278)$. }
\label{scenario-construction}
\end{figure}

For example, Figure \ref{scenario-construction} gives snap-shots of the sorting
procedure. Part (b)  shows the forest after the three first operations, the fourth
fission splits a tree with six nodes into two trees each with three nodes, 
corresponding to the cycle splitting of the fourth operation in the running example.

%
%

\section{Discussion and conclusions}
\label{Discussion}

In this paper, we presented results on the enumeration and representations of sorting
scenarios between co-tailed genomes. Since we introduced many combinatorial objects, 
we bypassed a lot of the usual material presented in rearrangement papers. The following topics
will be treated in a future paper.

The first topic is the complexity of the algorithms for switching between representations.
Algorithms 1 and 2 are not meant to be efficient, they are rather explicit descriptions of what 
is being computed. Preliminary work indicates that with suitable data structure, they can 
be implemented in $O(n)$ running time. Indeed, most of the needed information
can be obtained in a single traversal of a tree.

The second obvious extension is to generalize the enumeration formulas and representations
to arbitrary genomes. In the general case, when genomes are not necessarily co-tailed, the adjacency 
graph can be decomposed in cycles and paths, and additional sorting operations must be
considered, apart from operations that split cycles \cite{Bergeron-2006}. However, these new sorting
operations that act on paths create new paths that behave essentially like cycles.

We also had to defer to a further paper the details of the diverse uses of these new representations.
One of the main benefits of having a representation of a sorting scenario as a parking function, for example, is
that it solves the problem of uniform sampling of sorting scenarios \cite{Barcucci-1999}. There is no more bias attached
to choosing a first sorting operation, since, when using parking functions, the nature of the first
operation depends on the whole scenario. The representation of sorting scenarios as 
non-crossing partitions refinement also greatly helps in analyzing commutation and
conservation properties.

%
%
\bibliography{biblio}
\bibliographystyle{plain}

%
%

\end{document}